\documentclass[12pt]{article}

\usepackage{graphicx}
\usepackage{a4}
\usepackage{bm}

\def\div{\mathop{\rm div}}
\def\rot{\mathop{\rm rot}}

\begin{document}

\begin{titlepage}
\title{\hfill\parbox{4cm}
       {\normalsize UT-08-26\\August 2008}\\
       \vspace{1.5cm}
Quiver Chern-Simons theories and crystals
       \vspace{1.5cm}}
\author{
Yosuke Imamura\thanks{E-mail: \tt imamura@hep-th.phys.s.u-tokyo.ac.jp}\quad
and\quad
Keisuke Kimura\thanks{E-mail: \tt kimura@hep-th.phys.s.u-tokyo.ac.jp}
\\[20pt]
{\it Department of Physics, University of Tokyo,}\\ {\it Tokyo 113-0033, Japan}
}
\date{}

\maketitle
\thispagestyle{empty}

\vspace{0cm}

\begin{abstract}
\normalsize
We consider ${\cal N}=2$ quiver Chern-Simons theories
described by brane tilings,
whose moduli spaces are  toric Calabi-Yau $4$-folds.
Simple prescriptions to obtain
toric data of the moduli space
and a corresponding brane crystal
from a brane tiling
are proposed.
\end{abstract}

\end{titlepage}


\section{Introduction}

Recently, three-dimensional supersymmetric Chern-Simons theories
have attracted great interest as theories for multiple M2-branes in
various backgrounds.
This was triggered by the proposal of ${\cal N}=8$ interacting Chern-Simons
theories by Bagger, Lambert\cite{Bagger:2006sk,Bagger:2007jr,Bagger:2007vi},
and Gusstavson\cite{Gustavsson:2007vu,Gustavsson:2008dy}.
Their model (BLG model) is based on Lie $3$-algebra, and
the action includes structure constants, which satisfy the so-called
fundamental identity.
This model, however, has not succeeded in describing
an arbitrary number of M2-branes in uncompactified flat background,
due to the fact that the fundamental identity is
very restrictive and it admits the only one non-trivial finite-dimensional
algebra with a positive definite metric\cite{Papadopoulos:2008sk,Gauntlett:2008uf}.
The resulting theory is conjectured to describe two M2-branes on a certain orbifold\cite{Lambert:2008et,Distler:2008mk}.

Aharony et al. proposed alternative model in \cite{Aharony:2008ug},
based on the recent progress in ${\cal N}=4$ Chern-Simons theories\cite{Gaiotto:2008sd,Hosomichi:2008jd}.
Their model (ABJM model) is $U(N)\times U(N)$ Chern-Simons gauge theory at level $(k,-k)$
with bi-fundamental matter fields.
The model describes $N$ M2-branes in the ${\bf C}^4/{\bf Z}_k$
orbifold background.
Although only ${\cal N}=6$ supersymmetry is manifest in the model,
the supersymmetry is expected somehow to be enhanced to ${\cal N}=8$
when $k=1,2$.

After the proposal of the ABJM model,
some generalizations have been studied.
Orbifolds of the ABJM model are discussed in
\cite{Benna:2008zy,Imamura:2008nn,Terashima:2008ba}.
In \cite{Imamura:2008nn}, a certain class of ${\cal N}=3$
quiver Chern-Simons theories
with non-toric moduli spaces
are also studied based on the brane construction,
and the hyper-K\"ahler toric structure of the moduli spaces
is clarified in \cite{Jafferis:2008qz}.
The moduli spaces of other superconformal Chern-Simons theories
are studied in \cite{Fuji:2008yj,Hosomichi:2008jb,Kim:2008gn,Aharony:2008gk,Ooguri:2008dk}.

${\cal N}=2$ Chern-Simons theories with general quiver structure
are studied in \cite{Martelli:2008si},
and it is shown
how the gauge symmetries and D-term conditions are
modified compared to the case of four-dimensional ${\cal N}=1$
gauge theories described by the same quiver diagrams.
It is also found that the moduli spaces of such theories
generically include a baryonic branch.

In the case of four-dimensional ${\cal N}=1$ supersymmetric
gauge theories, brane tilings\cite{Hanany:2005ve,Franco:2005rj,Franco:2005sm}
are convenient tools to establish the relation between
gauge theories and their moduli spaces,
for a class of theories whose moduli spaces are toric Calabi-Yau
$3$-folds.
See \cite{Kennaway:2007tq,Yamazaki:2008bt} for review of brane tilings.
Brane tilings are expected to be convenient
for three-dimensional Chern-Simons theory, too.
${\cal N}=2$ quiver Chern-Simons theories described by
brane tilings are studied in \cite{Hanany:2008cd},
and it is shown that the moduli space of the theories are
toric Calabi-Yau $4$-folds,
and the Hilbert series is computed for some examples.

In this paper we consider
the class of ${\cal N}=2$ quiver Chern-Simons theories
described by brane tilings.
Our aim is to establish the relation between
brane tilings and brane crystals.
Brane crystals are three-dimensional graphs
proposed in \cite{Lee:2006hw,Lee:2007kv,Kim:2007ic}
as diagrams describing three-dimensional superconformal field
theories and the structure of their moduli spaces.
We first give a simple prescription to
obtain toric data of the moduli space
from a tiling,
and explain how we can construct a crystal
describing the same moduli space.

This paper is organized as follows.
In the next section, we explain the relation between
brane tilings and quiver Chern-Simons theories.
In section \ref{gaugesym.sec} we review how gauge symmetries
of Chern-Simons theories
are broken due to the existence of Chern-Simons terms
following \cite{Martelli:2008si}.
In section \ref{gio.sec}, we define
gauge invariant operators which parameterize the moduli spaces
of Chern-Simons theories.
In section \ref{toric.sec} we give a simple prescription
to obtain the toric data of the moduli spaces by using
tilings.
This section has some overlap with \cite{Ueda:2008hx}.
The relation between brane tilings and brane crystals
are discussed in \ref{crystal.sec}.
The last section is devoted to conclusions.

\section{Tilings and Chern-Simons theories}\label{tiling.sec}
We consider three-dimensional ${\cal N}=2$
quiver Chern-Simons theories
described by brane tilings, which are also studied in \cite{Hanany:2008cd}.

A brane tiling is a bipartite graph
drawn on ${\bf T}^2$.
A bipartite graph is a graph consisting of
vertices of two colors, say, white and black, and
all links connect two vertices with different colors.
Tilings have been used to describe four-dimensional
${\cal N}=1$ quiver gauge theories and the structure of
their moduli spaces.
The gauge group, the matter content, and the superpotential
of a gauge theory can be read off from the brane tiling for the theory.
Namely, faces correspond
to $U(N)$ factors in the gauge group,
and links to bi-fundamental fields.
The superpotential can be also read off from
the tiling in the way we will mention later.
These correspondences are naturally understood by regarding the tiling
as a NS5-D5 system in type IIB string theory.

\begin{figure}[htb]
\centerline{\includegraphics{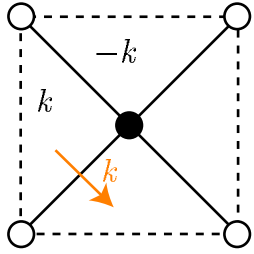}}
\caption{The tiling for the ABJM model at level $(k,-k)$.
The arrow represents the flow ${\bm s}$
defined in (\ref{ks}).}
\label{abjm.eps}
\end{figure}

In this paper, we use tilings to describe three-dimensional
${\cal N}=2$ Chern-Simons theories.
The gauge group, the matter content, and the superpotential
are read off from the tiling in the same way as the four-dimensional case.
These rules are naturally understood by regarding the
tiling as a D4-NS5 system in type IIA theory, rather than the type IIB
brane system.
By this reason, when we want to specify
which of three or four dimensional theory
a brane tiling describes,
we call it IIB tiling (for four-dimensional theory),
or IIA tiling (for Chern-Simons theory).
Figure \ref{abjm.eps} shows an example of IIA tiling for the ABJM model.

Because we are here interested in the structure of
the background spacetime probed by M2-branes,
we discuss only abelian ($N=1$) case.
We use indices $I,J,\ldots$ for links,
$i,j,\ldots$ for faces, and $a,b,\ldots$ for vertices.
Let $U(1)_i$ be the gauge group associated with face $i$,
and $\Phi_I$ be the bi-fundamental field associated with link $I$.
$I\in i$ means the link $I$ is on the face $i$.
The bi-fundamental field $\Phi_I$
is charged under two $U(1)$ factors corresponding to the
two faces sharing the link.
The $U(1)_i$ charge $Q_{Ii}$ of the chiral multiplet $\Phi_I$ is uniquely
determined by the bipartite graph.
When the link $I$ is not a side of the face $i$ $Q_{Ii}=0$.
$Q_{Ii}$ is $+1$ ($-1$)
if the left endpoint of the link $I$ is black (white) when
it is seen from the face $i$.

In order to specify a Chern-Simons theory,
we need to fix the levels $k_i\in{\bf Z}$ for each gauge group
as numbers assigned to faces in a IIA tiling.
As is pointed out in \cite{Martelli:2008si},
we need to impose the condition
\begin{equation}
\sum_i k_i=0,
\label{total0}
\end{equation}
to obtain a four-dimensional
moduli space.
Because of this condition
we can represent the levels $k_i$ as
\begin{equation}
k_i=\sum_I Q_{Ii} s_I.
\label{ks}
\end{equation}

\begin{figure}[htb]
\centerline{\includegraphics{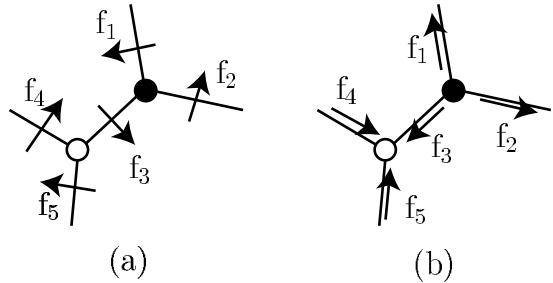}}
\caption{For a set of numbers $f_I$ assigned to links
we define two flows.
(a) is a normal flow $\bm f$ and (b) is a tangential flow $\bm f^*$.
These two flows are related by the $\pi/2$ rotation of arrows.}
\label{flows.eps}
\end{figure}

In order to interpret relations like (\ref{ks}) geometrically,
we define two kinds of flows on the tiling.
Let $f_I$ be a set of numbers assigned to links.
A normal flow $\bm f$ is a flow from faces to faces.
We define the orientation of the flow to be
the anti-clockwise direction around
black vertices. ((a) in Figure \ref{flows.eps})
The other flow associated with $f_I$ is the tangential flow ${\bm f}^*$,
which describes flow along links from black vertices to white ones.
 ((b) in Figure \ref{flows.eps})
 We can rewrite
the relation (\ref{ks})
in terms of the
normal flow ${\bm s}$ or
the tangential flow ${\bm s}^*$
as
\begin{equation}
\{k_i\}=\div{\bm s}=\rot{\bm s}^*.
\label{divrot}
\end{equation}
We will later see that this relation has natural interpretation
in the context of brane realization of Chern-Simons theories.

\begin{figure}[htb]
\centerline{\includegraphics{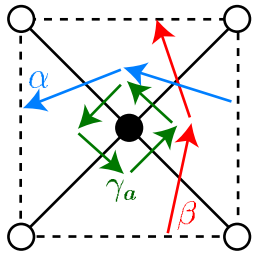}}
\caption{Examples of cycles $\bm\alpha$, $\bm\beta$ and ${\bm\gamma}_a$ for the ABJM tiling are shown.}
\label{abg.eps}
\end{figure}

For later convenience, we introduce the following
normal flows. (See Figure \ref{abg.eps}.)
\begin{itemize}
\item ${\bm\alpha}$ : a unit flow along $\alpha$-cycle on the torus.
\item ${\bm\beta}$ : a unit flow along $\beta$-cycle on the torus.
\item ${\bm\gamma}_a$ : a unit flow around vertex $a$.
The orientation is anti-clockwise (clockwise) around black (white) vertices.
\end{itemize}
An arbitrary conserved flow can be given as a linear combination
of these flows.
We define an operator ${\cal O}_{\bm f}$ for a normal flux $\bm f$ with
non-negative integral components $f_I$ by
\begin{equation}
{\cal O}_{\bm f}=\prod_I\Phi_I^{f_I}.
\label{ofdef}
\end{equation}
If ${\bm f}$ is conserved flow satisfying $\div{\bm f}=0$,
the operator is also defined for general $N$
as a single or multiple trace operator, which is
often called mesonic operators.
Baryonic operators are associated with non-conserved flows.
With this notation, the superpotential is represented as
\begin{equation}
W=\sum_a\pm{\cal O}_{{\bm \gamma}_a},
\label{superpot}
\end{equation}
where the signature of the summand is positive (negative)
for  black (white) vertices.

\section{Gauge symmetries}\label{gaugesym.sec}
In both three- and four-dimensional cases,
the moduli space is defined as the coset
$X/G$, where $X$ is the manifold defined by the F-term conditions
and $G$ is the complexified gauge group.
Because IIA and IIB tilings give the same F-term conditions,
the manifold $X$ is common to two cases.
A difference arises in the gauge symmetry $G$.
In this section we review how this
difference arises following \cite{Martelli:2008si}.

If the tiling has $n$ faces, there are $n$ $U(1)$ factors.
Among them, the diagonal $U(1)$ decouples from matter fields,
and the effective gauge symmetry is $U(1)^{n-1}$.
The complexification of this symmetry gives $G$ in the IIB case.
In the IIA case, however, it is known that $U(1)^{n-1}$ is
broken down to $U(1)^{n-2}$ due to
the existence of the Chern-Simons terms.

Let $A_i$ be the $U(1)_i$ gauge field.
We define gauge fields
\begin{equation}
a=\sum_{i=1}^nA_i,\quad
b=\sum_{i=1}^nk_iA_i.
\end{equation}
Let $c_k$ ($k=1,\ldots,n-2$) be linear comminations of
$A_i$ linearly independent of
$a$ and $b$.
We can rewrite the Chern-Simons terms in the form
\begin{equation}
S_{\rm CS}=\frac{1}{2\pi n}\int b\wedge f+S'[b,c_k]
\label{scs}
\end{equation}
where $f=da$ and $S'$ does not depend on the diagonal $U(1)$ gauge field $a$.
Because the gauge field $a$ does not couple to matter fields,
it appears only in the first term of (\ref{scs}).
The action includes $a$ only through $f$,
and we can dualize it by introducing Lagrange multiplier $\tau$
and adding the following term to the action.
\begin{equation}
S_\tau=-\frac{1}{2\pi}\int d\tau\wedge f.
\end{equation}
The equation of motion for $f$ is
\begin{equation}
d\tau=\frac{1}{n}b.
\label{dtaub}
\end{equation}

Let us consider gauge transformation
\begin{equation}
A_i\rightarrow A_i+d\theta_i.
\label{gaugetr}
\end{equation}
The relation (\ref{dtaub})
implies that the dual scalar field $\tau$
should be transformed
under (\ref{gaugetr}) by
\begin{equation}
\delta\tau=\frac{1}{n}\sum_{i=1}^nk_i\theta_i.
\end{equation}
This non-linear gauge transformation of $\tau$ means that
the gauge symmetry is always partially broken due to the vev
of the scalar field $\tau$.
As is shown in \cite{Martelli:2008si}
the period of $\tau$ is $2\pi/n$,
and the parameters for unbroken gauge transformations
should satisfy
\begin{equation}
2\pi{\bf Z}
\ni\sum_{i=1}^nk_i\theta_i
=\sum_{I,i}Q_{Ii} s_I\theta_i,
\label{paramconst}
\end{equation}
where we used (\ref{ks}) to obtain the final expression.

\section{Gauge invariant operators}\label{gio.sec}
When we analyze the moduli space of a gauge theory,
it is convenient to use gauge invariant operators
as coordinates of the moduli space.
In four-dimensional gauge theories described by
IIB brane tiling, it is known that
the moduli space is parameterized by
three gauge invariant operators
\begin{equation}
{\cal M}_\alpha={\cal O}_{\bm\alpha},\quad
{\cal M}_\beta={\cal O}_{\bm\beta},\quad
{\cal W}={\cal O}_{{\bm\gamma}_a},
\label{mesonic}
\end{equation}
associated with the flows defined in section \ref{tiling.sec}.
For general $N$, these operators are defined as single-trace mesonic
operators.
${\cal W}$ is one of the terms in the superpotential (\ref{superpot}).
Due to the $F$-term conditions, all terms in the superpotential
have the same vev, and ${\cal W}$ as an element of the chiral ring
does not depend on the choice of the vertex $a$.
Because $\bm\alpha$, $\bm\beta$, and ${\bm\gamma}_a$ generate
arbitrary conserved flows,
an arbitrary mesonic operator can be written as a function
of these mesonic operators, and we can use these three as
coordinates in the three-dimensional moduli space of
four-dimensional gauge theory.

In the case of Chern-Simons theory,
(\ref{mesonic}) are again gauge invariant operators,
and we can use them as coordinates in the moduli space.
However, we need another gauge invariant operator to parameterize
the four-dimensional moduli space.
Indeed, the restriction (\ref{paramconst})
of the gauge transformation parameters admits
extra gauge invariant operators in addition to
the mesonic operators in the four-dimensional gauge theory.

Let us consider an operator ${\cal O}_{\bm q}$ associated with
a flow ${\bm q}$, which is not necessarily conserved.
The field $\Phi_I$ associated with link $I$
is transformed under the gauge transformation (\ref{gaugetr}) as
\begin{equation}
\Phi_I\rightarrow
\exp\left(i\sum_iQ_{Ii}\theta_i\right)\Phi_I,
\end{equation}
and
the gauge transformation of the operator ${\cal O}_{\bm q}$ is
\begin{equation}
{\cal O}_{\bm q}\rightarrow
\exp\left(i\sum_{I,i}Q_{Ii}q_I\theta_i\right)
{\cal O}_{\bm q}.
\label{oq}
\end{equation}
For the operator to be gauge invariant,
the components $q_I$ of the flow must satisfy
\begin{equation}
2\pi{\bf Z}
\ni\sum_{I,i}Q_{Ii} q_I\theta_i.
\end{equation}
If this condition were imposed for arbitrary $\theta_i$,
solutions would be given by $q_I=c_I$ where $c_I$ is an arbitrary
flow satisfying
\begin{equation}
\sum_IQ_{Ii}c_I=0\quad\forall\ i.
\label{ci}
\end{equation}
This is equivalent to $\div{\bm c}=0$, and the normal flow ${\bm c}$ is
conserved.
Solutions in this form correspond to mesonic operators
generated by (\ref{mesonic}).

The parameters $\theta_i$ are, however, constrained by (\ref{paramconst})
in the Chern-Simons theory.
Thus we have an extra solution
$q_I=s_I$,
and a general solution is given by
\begin{equation}
q_I=ms_I+c_I,
\end{equation}
where $m$ is an arbitrary integer and $c_I$ is
a conserved flow satisfying (\ref{ci}).
Therefore, as the fourth coordinate
on the moduli space, we should introduce
the following baryonic operator associated with ${\bm s}$.
\begin{equation}
{\cal B}={\cal O}[{\bm s}]=\prod_I \Phi_I^{s_I}.
\label{baryonic}
\end{equation}
An arbitrary gauge invariant operator
in the Chern-Simons theory is given as a function of
the four operators
\begin{equation}
{\cal M}_\alpha,\  {\cal M}_\beta,\   {\cal B},\ {\cal W}.
\label{fourops}
\end{equation}

\section{Toric data}\label{toric.sec}
In this section, we give a simple prescription
to obtain toric data of the moduli space of Chern-Simons theory
from the IIA brane tiling for the theory.
The same subject is also investigated in \cite{Ueda:2008hx}.
 
In general a toric Calabi-Yau $n$-fold is represented as
a ${\bf T}^n$ fibration over an $n$-dimensional polyhedral cone ${\cal C}$.
The boundary of ${\cal C}$ consists of $(n-1)$-fans.
On each $(n-1)$-fan a cycle $v$ in the toric fiber shrinks.
In other words, the fan is the fixed submanifold of the $U(1)$ isometry
generated by the vector $v$.
For each $(n-1)$-fan, there is a vector $v$ representing the shrinking cycle,
and the toric data is given as a set of such vectors.

In order to extract the toric data of the moduli space from
the information of a gauge theory,
it is convenient to translate the system into a gauged linear sigma model (GLSM).
This is achieved by solving the F-term conditions with the help of perfect matchings.

\begin{figure}[htb]
\centerline{\includegraphics{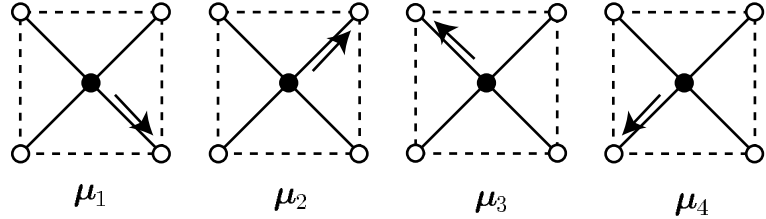}}
\caption{The four perfect matchings of the ABJM tiling.}
\label{pm.eps}
\end{figure}

A perfect matching is a number assignment $\mu_I$ to links in a tiling
which satisfies the following conditions.
\begin{itemize}
\item
$\mu_I=0$ or $1$ for any link $I$.
\item
Among links ending on a vertex $a$, only one has non-vanishing $f_I$.
\end{itemize}
The following equation follows from these two conditions.
\begin{equation}
\langle{\bm\gamma}_a,{\bm\mu}^*\rangle
\equiv \sum_{I\in a}f_I=1
\quad
\forall a,
\label{sumisone}
\end{equation}
${\bm\mu}^*$ is the tangential flow associated with
the number assignment $\mu_I$, and
the product $\langle *,*\rangle$ is the intersection of
a normal flow and a tangential flow, which is defined by
\begin{equation}
\langle{\bm f},{\bm g}^*\rangle=\sum_I f_Ig_I.
\end{equation}
Figure \ref{pm.eps} shows the four perfect matchings of
the ABJM tiling.

The F-term conditions require all the terms ${\cal O}_{{\bm\gamma}_a}$ in the superpotential are the same.
We can solve this condition by\cite{Franco:2006gc}
\begin{equation}
\Phi_I=\prod_{\mu\ni I}\rho_\mu,
\label{fcsol}
\end{equation}
where $\rho_\mu$ is a GLSM field defined for each perfect matching $\mu$,
and the summation is taken over all the perfect matchings
with $\mu_I=1$.
Substituting this into the definition of ${\cal O}_{{\bm\gamma}_a}$ and
using (\ref{sumisone}), we can show that ${\cal O}_{{\bm\gamma}_a}$
is the product of all the GLSM fields regardless of the index $a$,
and the $F$-term conditions are therefore satisfied
by (\ref{fcsol}).
Because this expression is redundant, we need to extend
the gauge symmetry $G$ acting on $\Phi_I$
by adding $U(1)$ rotations of GLSM fields which keep $\Phi_I$ invariant.
Let $G'$ be this extended gauge symmetry.
If the number of the perfect matchings is $n_{\rm pm}$ and
the space spanned by the GLSM fields is ${\bf C}^{n_{\rm pm}}$,
the moduli space of the gauge theory is
the coset ${\bf C}^{n_{\rm pm}}/G'$.

In order to obtain the toric data,
we need to find $U(1)$ symmetries which have
non-trivial fixed submanifolds.
It is easy to show that in the moduli space defined as the
coset ${\bf C}^{n_{\rm pm}}/G'$,
such a submanifold is given as the image of
the fixed plane $\rho_\mu=0$ of $U(1)_\mu$ symmetry
by the homomorphism ${\bf C}^{n_{\rm pm}}\rightarrow{\bf C}^{n_{\rm pm}}/G'$,
where
$U(1)_\mu$ is the symmetry which rotate only one GLSM field
$\rho_\mu$ with charge $1$.
Thus, the components of the killing vector $v_\mu$
are given as the $U(1)_\mu$ charges of the four toric coordinates.

As we mentioned above,
we can use
the four gauge invariant operators in (\ref{fourops})
as coordinates in the Calabi-Yau $4$-fold,
and then the four components of $v_\mu$
are $U(1)_\mu$ charges of these operators.
By substituting
(\ref{fcsol}) into (\ref{ofdef})
we rewrite an operator ${\cal O}_{\bm q}$ in terms of
GLSM fields as
\begin{equation}
{\cal O}_{\bm q}
=\prod_I\prod_{\mu\ni I}\rho_\mu^{q_I}
=\prod_\mu\rho_\mu^{\langle{\bm q},{\bm\mu}^*\rangle},
\end{equation}
and thus, the $U(1)_\mu$ charge $[{\cal O}_{\bm q}]_\mu$ of
the operator ${\cal O}_{\bm q}$ is given by
\begin{equation}
[{\cal O}_{\bm q}]_\mu=\langle{\bm q},{\bm\mu}^*\rangle.
\end{equation}
Applying this formula to the four gauge invariant operators,
we obtain the following components of the killing vectors
\begin{equation}
v_\mu
=\left(\begin{array}{c}
{}[{\cal M}_\alpha]_\mu \\
{}[{\cal M}_\beta]_\mu \\
{}[{\cal B}]_\mu \\
{}[{\cal W}]_\mu
\end{array}\right)
=\left(\begin{array}{c}
\langle {\bm\alpha},{\bm\mu}^*\rangle \\
\langle {\bm\beta},{\bm\mu}^*\rangle \\
\langle {\bm s},{\bm\mu}^*\rangle \\
1
\end{array}\right)
\label{toric4d}
\end{equation}
The last components of these vectors are always $1$, and this guarantees that the
toric manifold is Calabi-Yau.
With this formula, we can easily obtain toric data
from a given IIA tiling.

\begin{figure}[htb]
\centerline{\includegraphics{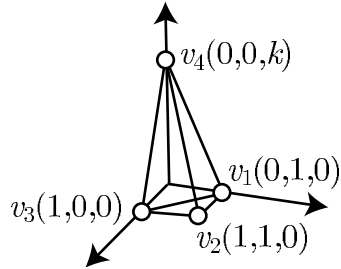}}
\caption{The toric diagram of the orbifold ${\bf C}^4/{\bf Z}_k$.}
\label{c4zk.eps}
\end{figure}

As a simple example, let us consider the ABJM tiling
in Figure \ref{abjm.eps}.
If we use the four perfect matchings in Figure \ref{pm.eps},
and flows ${\bm s}$, $\bm\alpha$, $\bm\beta$, and ${\bm\gamma}_a$
in Figure \ref{abjm.eps} and \ref{abg.eps},
we obtain the following four killing vectors.
\begin{equation}
v_1=\left(\begin{array}{c}
0 \\ 1 \\ 0 \\ 1
\end{array}\right),\quad
v_2=\left(\begin{array}{c}
1 \\ 1 \\ 0 \\ 1
\end{array}\right),\quad
v_3=\left(\begin{array}{c}
1 \\ 0 \\ 0 \\ 1
\end{array}\right),\quad
v_4=\left(\begin{array}{c}
0 \\ 0 \\ k \\ 1
\end{array}\right).
\end{equation}
By neglecting the fourth components of these vectors
and plotting corresponding points in the three-dimensional lattice,
we obtain the toric diagram of the moduli space ${\bf C}^4/{\bf Z}_k$.
(Figure \ref{c4zk.eps})

\section{Relation to crystals}\label{crystal.sec}
A brane tiling describing a four-dimensional gauge theory
can be regarded as a brane systems consisting of D5-branes and NS5-branes,
which is T-dual to D3-branes probing a toric Calabi-Yau $3$-fold,
and the rules of reading off the gauge theory from the tiling
have natural interpretation in terms of this brane system.
For example, faces in a brane tiling represent
D5-branes in the brane system,
and the $U(N)$ factors in the gauge group
are identified with the gauge groups realized on the D5-branes.

Brane crystals\cite{Lee:2006hw,Lee:2007kv,Kim:2007ic}
are analogues of brane tilings for M2-branes probing
four-dimensional toric CY cones.
By T-duality transformation in M-theory,
a system of M2-branes probing a four-dimensional toric Calabi-Yau cone
is transformed into a brane system consisting of M5-branes.
Brane crystals are bipartite graphs in ${\bf T}^3$ representing
the structure of the M5-brane systems\cite{Lee:2006hw}.

Contrary to the case of brane tilings for four-dimensional gauge theories,
we can obtain much less information from this brane system.
This is because we have only little knowledge about theories realized on
M5-brane systems.
The purpose of this section is to obtain some information
about the relation between Chern-Simons theories and brane crystals
by using the results obtained in the previous sections.

\begin{figure}[tb]
\centering
\includegraphics[width=10cm]{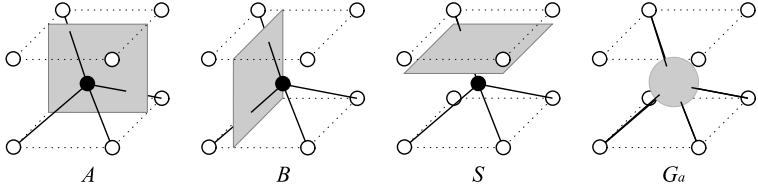}
\caption{2-cycles used to extract toric data from a crystal
are shown.
These are expected to represent M2-branes corresponding to
gauge invariant operators\cite{Lee:2006hw,Lee:2007kv}.}
\label{2-cycles}
\end{figure}

The method to obtain toric data from crystals
has been already known\cite{Lee:2007kv}.
Actually, we may define brane crystals
as bipartite graphs in ${\bf T}^3$ 
which give toric data of Calabi-Yau $4$-folds
in a similar way as brane tilings.
The method to obtain the toric diagram from a brane crystal is as follows:
First, instead of the $\alpha$ and $\beta$ cycles in brane tilings,
we define three closed 2-cycles $A$ ($1$-$3$ plane),
$B$ ($2$-$3$ plane), and $S$ ($1$-$2$ plane).
(See Figure \ref{2-cycles}.)
We assume these do not include vertices of the crystal on them.
We also define a cycle $G_a$ for each vertex $a$, which is a sphere enclosing the vertex $a$.
(See Figure \ref{2-cycles}.)
Because the crystal is bipartite as well as tilings,
we can define perfect matchings on it.
If we denote the intersection number
of a $2$-cycle $C$ and a perfect matching $\mu$ by
$[C,\mu]$, the vectors $v_\mu$ forming the toric diagram
are given by
\begin{equation}
v_\mu
=\left(\begin{array}{c}
[A,\mu] \\
{}[B,\mu] \\
{}[S,\mu] \\
{}[G_a,\mu]
\end{array}\right).
\label{crystalformula}
\end{equation}
By definition, the last component $[G_a,\mu]$ is always $1$,
and the Calabi-Yau condition is satisfied.

\begin{figure}[tb]
\centering
\begin{tabular}{cc}
\includegraphics[width=30mm]{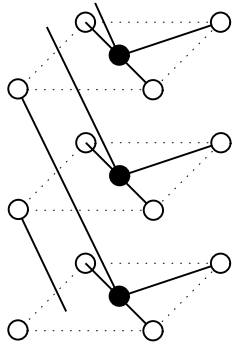}
\end{tabular}
\caption{A crystal for ABJM model with $k=2$ is shown.
This figure includes two fundamental regions.}
\label{tiling_crystal}
\end{figure}

When a IIA tiling is given,
it is easy to construct a brane crystal
which gives the same toric data
by the formula (\ref{crystalformula})
as the data obtained by (\ref{toric4d}) from the IIA tiling.
Let $(x_a,y_a)$ be the coordinates
of the vertex $a$ in the IIA tiling.
We put the corresponding vertex in the crystal
at the point $(x_a,y_a,0)$ in the three-dimensional torus.
We make the three-dimensional graph by connecting these
vertices in the same way as the tiling.
Namely, if vertices $a$ and $b$ in the tiling are connected by a link,
we connect the corresponding points in ${\bf T}^3$ by a link, too.
There are infinitely many ways of connecting two vertices
in ${\bf T}^3$ with a link with different winding numbers.
We fix this ambiguity by requiring the following two conditions.
\begin{itemize}
\item The crystal reduces to the original tiling by the projection
along the vertical axis.
\item The vertical winding number of link $I$ is $s_I$.
\end{itemize}
In other words, we interpret the integers $s_I$ assigned to links
as the gradient of links in the three-dimensional space.
See Figure \ref{tiling_crystal} for an example of the ABJM model with $k=2$.
As the result, we obtain a three-dimensional bipartite graph
with the same number of vertices and links as the original brane tiling
(Figure~\ref{tiling_crystal}).

Let us confirm that the crystal
constructed in this way correctly reproduces the
toric data (\ref{toric4d}) obtained in the previous section.
First of all, the two-dimensional and three-dimensional graphs
are differ only by their embeddings
to the tori.
The former is embedded in ${\bf T}^2$ and the latter is in ${\bf T}^3$.
Therefore, the three-dimensional graph has the same perfect matchings
as the two-dimensional one.

Let us first consider the first two components
$\langle{\bm\alpha},{\bm\mu}^*\rangle$ and
$\langle{\bm\beta},{\bm\mu}^*\rangle$ in (\ref{toric4d}).
We define the three-dimensional lift
of $\alpha$ and $\beta$ cycles by
\begin{equation}
A=\alpha\otimes {\bf S}^1,\quad
B=\beta\otimes {\bf S}^1,
\end{equation}
where ${\bf S}^1$ here is the cycle along the vertical direction.
We use these $2$-cycles as $A$ and $B$ in the formula (\ref{crystalformula}).
Then it is obvious that the first two components of (\ref{toric4d})
and those in (\ref{crystalformula}) are the same.
\begin{equation}
\langle{\bm\alpha},{\bm\mu}^*\rangle=[ A,\mu],\quad
\langle{\bm\beta},{\bm\mu}^*\rangle=[ B,\mu].
\end{equation}

For the last components,
we define 2-cycle $\gamma_a\otimes{\bf S}^1$ for each vertex $a$.
These are homologous to $G_a$ defined above,
and the relation
\begin{equation}
\langle{\bm\gamma}_a,{\bm\mu}^*\rangle=[ G_a,\mu]
\end{equation}
holds.
(Actually, this is by definition always $1$.)

Finally, let us consider the third component in (\ref{toric4d}),
which is given as the $U(1)_\mu$ charge of the baryonic operator ${\cal B}$.
In the brane tiling,
the baryonic operator is expressed in different way from the other
mesonic operators.
Mesonic operators are associated with conserved flows on the tiling,
while the flow ${\bm s}$ corresponding to the baryonic operator ${\cal B}$
is not conserved.
In the crystal, however,
the third component is also given as
the intersection of closed 2-cycle and perfect matchings.
As we mentioned above, the link $I$ in the crystal
has the vertical winding $s_I$,
and it intersects $s_I$ times with the 2-cycle $S$.
Therefore, the intersection $\langle{\bm s},{\bm\mu}^*\rangle$
can be rewritten as the intersection of the closed $2$-cycle $S$ and
the perfect matching $\mu$.
\begin{equation}
\langle{\bm s},{\bm\mu}^*\rangle
=[S,\mu ].
\end{equation}
Thus, the third components of (\ref{toric4d}) and (\ref{crystalformula})
are the same.

Now we have
confirmed that the crystal constructed above correctly reproduces
the toric data of the moduli space of the Chern-Simons theory
described by the tiling.
At the same time, we have
established the correspondence between
the gauge invariant operators in (\ref{fourops})
and closed 2-cycles in the crystal.
Because the set of operators in (\ref{mesonic}) and (\ref{baryonic})
generates arbitrary
gauge invariant operators,
we have established the complete map between gauge invariant operators
including both mesonic and baryonic ones and closed $2$-cycle in the crystal,
which are interpreted as closed M2-branes\cite{Lee:2006hw,Lee:2007kv}.
An interesting feature of this correspondence
is that even though in the Chern-Simons theory
baryonic operators and mesonic operators have
different structure,
the brane crystal describes these in
the parallel way.

As another support to our prescription,
we can show that the level $k_I$ are naturally obtained
from the brane system described by the crystal.
In order to read off the Chern-Simons theory from a
crystal, we need to project the crystal along
the vertical direction, and go back to the tiling.
From the viewpoint of brane system,
we can interpret this projection as
the compactification of M-theory to type IIA string theory.
Then, links and faces in the tiling are interpreted as
a network of NS5-brane and D4-branes ending on the NS5-brane, respectively.
Gauge groups are realized on the D4-branes,
and the Chern-Simons terms are induced from
the following boundary term in the D4-brane action.
\begin{equation}
S=\frac{1}{2\pi}\int_{\partial{\rm D4}} A\wedge dA\wedge d\phi,
\end{equation}
where $A$ is the gauge field on the D4-brane
and $\phi$ is the compact scalar field on the NS5-brane
corresponding to the $X^{11}$ coordinate in the M-theory picture.
If the scalar field $\phi$ has non-trivial profile
along the boundary of the D4-brane,
this induces the Chern-Simons coupling in the three-dimensional
gauge theory,
and the level is given by
\begin{equation}
k_i=\oint d\phi,
\label{kintdp}
\end{equation}
where the integration is taken over the boundary of the face $i$.
If we identify $d\phi$ as the gradient $s_I$ along links,
(\ref{kintdp}) is nothing but the relation (\ref{ks}), or, equivalently,
(\ref{divrot}).

\section{Conclusions}
In this paper we investigated the relation
between brane tilings describing ${\cal N}=2$ Chern-Simons
theories and the toric data of their moduli spaces.
We gave a simple procedure to read off the toric data of the
moduli space from the brane tiling.
In order to obtain the toric data,
we should first represent the Chern-Simons levels as a flow ${\bm s}$
on the tiling, and the vectors $v_\mu$ forming the toric diagram
are obtained as the intersection of the perfect matchings ${\bm\mu}^*$
and the flows $({\bm\alpha},{\bm\beta},{\bm s})$.

IIA brane tilings, which are regarded as brane systems consisting
of D4-branes and NS5-branes, can be regarded as the projection
of the crystals, which describe M5-brane systems.
We showed that we can lift a IIA tiling to the corresponding crystal
by using the flow ${\bm s}^*$
as the gradient of links.
We found that gauge invariant operators, which include both
mesonic and baryonic ones,
are represented in the crystal as closed $2$-cycles.

We emphasize that
although our prescription
always gives a crystal
for a given IIA tiling,
it is not always possible
to give a tiling which reproduce
a given crystal.
Our prescription does not guarantee
the existence of a Chern-Simons theory
which reproduces a given toric Calabi-Yau $4$-fold
as its moduli space.
There may not exist corresponding Chern-Simons theories
for a class of manifolds.
Contrary, there are crystals which gives more than two tilings
by the projection along different directions.
This may suggest the duality among Chern-Simons theories.
We wish to come back to these issues in near future.

\section*{Acknowledgements}
We would like to thank M.~Yamazaki for valuable discussions.
Y.~I. is partially supported by
Grant-in-Aid for Young Scientists (B) (\#19740122) from the Japan
Ministry of Education, Culture, Sports,
Science and Technology.



\begin{thebibliography}{99}
\bibitem{Bagger:2006sk}
  J.~Bagger and N.~Lambert,
  Phys.\ Rev.\  D {\bf 75}, 045020 (2007)
  [arXiv:hep-th/0611108].

\bibitem{Bagger:2007jr}
  J.~Bagger and N.~Lambert,
  Phys.\ Rev.\  D {\bf 77}, 065008 (2008)
  [arXiv:0711.0955 [hep-th]].

\bibitem{Bagger:2007vi}
  J.~Bagger and N.~Lambert,
  JHEP {\bf 0802}, 105 (2008)
  [arXiv:0712.3738 [hep-th]].


\bibitem{Gustavsson:2007vu}
  A.~Gustavsson,
  arXiv:0709.1260 [hep-th].


\bibitem{Gustavsson:2008dy}
  A.~Gustavsson,
  JHEP {\bf 0804}, 083 (2008)
  [arXiv:0802.3456 [hep-th]].



\bibitem{Papadopoulos:2008sk}
  G.~Papadopoulos,
  JHEP {\bf 0805}, 054 (2008)
  [arXiv:0804.2662 [hep-th]].

\bibitem{Gauntlett:2008uf}
  J.~P.~Gauntlett and J.~B.~Gutowski,
  arXiv:0804.3078 [hep-th].


\bibitem{Lambert:2008et}
  N.~Lambert and D.~Tong,
  Phys.\ Rev.\ Lett.\  {\bf 101}, 041602 (2008)
  [arXiv:0804.1114 [hep-th]].


\bibitem{Distler:2008mk}
  J.~Distler, S.~Mukhi, C.~Papageorgakis and M.~Van Raamsdonk,
  JHEP {\bf 0805}, 038 (2008)
  [arXiv:0804.1256 [hep-th]].



\bibitem{Aharony:2008ug}
  O.~Aharony, O.~Bergman, D.~L.~Jafferis and J.~Maldacena,
  arXiv:0806.1218 [hep-th].

\bibitem{Gaiotto:2008sd}
  D.~Gaiotto and E.~Witten,
  arXiv:0804.2907 [hep-th].

\bibitem{Hosomichi:2008jd}
  K.~Hosomichi, K.~M.~Lee, S.~Lee, S.~Lee and J.~Park,
  JHEP {\bf 0807}, 091 (2008)
  [arXiv:0805.3662 [hep-th]].


\bibitem{Benna:2008zy}
  M.~Benna, I.~Klebanov, T.~Klose and M.~Smedback,
  arXiv:0806.1519 [hep-th].
\bibitem{Imamura:2008nn}
  Y.~Imamura and K.~Kimura,
  arXiv:0806.3727 [hep-th].
\bibitem{Terashima:2008ba}
  S.~Terashima and F.~Yagi,
  arXiv:0807.0368 [hep-th].


\bibitem{Jafferis:2008qz}
  D.~L.~Jafferis and A.~Tomasiello,
  arXiv:0808.0864 [hep-th].




\bibitem{Fuji:2008yj}
  H.~Fuji, S.~Terashima and M.~Yamazaki,
  arXiv:0805.1997 [hep-th].

\bibitem{Hosomichi:2008jb}
  K.~Hosomichi, K.~M.~Lee, S.~Lee, S.~Lee and J.~Park,
  arXiv:0806.4977 [hep-th].

\bibitem{Kim:2008gn}
  N.~Kim,
  arXiv:0807.1349 [hep-th].

\bibitem{Aharony:2008gk}
  O.~Aharony, O.~Bergman and D.~L.~Jafferis,
  arXiv:0807.4924 [hep-th].

\bibitem{Ooguri:2008dk}
  H.~Ooguri and C.~S.~Park,
  arXiv:0808.0500 [hep-th].




\bibitem{Martelli:2008si}
  D.~Martelli and J.~Sparks,
  arXiv:0808.0912 [hep-th].


\bibitem{Hanany:2005ve}
  A.~Hanany and K.~D.~Kennaway,
  arXiv:hep-th/0503149.
\bibitem{Franco:2005rj}
  S.~Franco, A.~Hanany, K.~D.~Kennaway, D.~Vegh and B.~Wecht,
  JHEP {\bf 0601} (2006) 096,
  arXiv:hep-th/0504110.
\bibitem{Franco:2005sm}
  S.~Franco, A.~Hanany, D.~Martelli, J.~Sparks, D.~Vegh and B.~Wecht,
  JHEP {\bf 0601} (2006) 128,
  arXiv:hep-th/0505211.

\bibitem{Kennaway:2007tq}
  K.~D.~Kennaway,
  Int.\ J.\ Mod.\ Phys.\  A {\bf 22}, 2977 (2007)
  [arXiv:0706.1660 [hep-th]].

\bibitem{Yamazaki:2008bt}
  M.~Yamazaki,
  arXiv:0803.4474 [hep-th].




\bibitem{Hanany:2008cd}
  A.~Hanany and A.~Zaffaroni,
  arXiv:0808.1244 [hep-th].

\bibitem{Lee:2006hw}
  S.~Lee,
  Phys.\ Rev.\  D {\bf 75}, 101901 (2007)
  [arXiv:hep-th/0610204].
\bibitem{Lee:2007kv}
  S.~Lee, S.~Lee and J.~Park,
  JHEP {\bf 0705}, 004 (2007)
  [arXiv:hep-th/0702120].

\bibitem{Kim:2007ic}
  S.~Kim, S.~Lee, S.~Lee and J.~Park,
  Nucl.\ Phys.\  B {\bf 797}, 340 (2008)
  [arXiv:0705.3540 [hep-th]].





\bibitem{Ueda:2008hx}
  K.~Ueda and M.~Yamazaki,
  arXiv:0808.3768 [hep-th].

\bibitem{Franco:2006gc}
  S.~Franco and D.~Vegh,
  JHEP {\bf 0611}, 054 (2006)
  [arXiv:hep-th/0601063].



\end{thebibliography}
\end{document}